\documentclass{jfm}
\usepackage{graphicx}
\usepackage{amsmath}

\shorttitle{Deformation characteristics of a single droplet driven by a piezoelectric nozzle of the drop-on-demand inkjet system}
\shortauthor{S. Wang, Y. Zhong, H. Fang}

\title{Deformation characteristics of a single droplet driven by a piezoelectric nozzle of the drop-on-demand inkjet system}

\author{Shangkun Wang\aff{1}, Yonghong Zhong\aff{1} and Haisheng Fang\aff{1}\corresp{\email{hafang@hust.edu.cn (Shangkun Wang and Yonghong Zhong share equal contribution to this paper)}}}

\affiliation{\aff{1}State Key Laboratory of Coal Combustion, School of Energy and Power Engineering, Huazhong University of Science and Technology, Wuhan, Hubei, 430074, China}
\begin{document}

\maketitle

\begin{abstract}
In the drop-on-demand (DOD) inkjet system, deformation process and the direct relations between the droplet motions and the liquid properties have been seldom investigated, although they are very critical for the printing accuracy. In this study, experiments and computational simulation regarding deformation of a single droplet driven by a piezoelectric nozzle have been conducted to address the deformation characteristics of  droplets. It is found that the droplet deformation is influenced by the pressure wave propagation in the ink channel related to the driven parameters and reflected in the subsequent droplet motions. The deformation extent oscillates with a certain period of T and a decreasing amplitude as the droplet moves downwards. The deformation extent is found strongly dependent on the capillary number (\textit{Ca}), first ascended and then descended as the number increases. The maximum value of the deformation extent is surprisingly found to be within range of 0.68-0.82 of the \textit{Ca} number regardless of other factors. Furthermore, the Rayleigh’s linear relation of the oscillation frequency of the droplet to the parameter, $\sqrt{\sigma /\rho {{r}^{3}}}$, is updated with a smaller slope shown both by experiments and simulation.
\end{abstract}


\section{Introduction}
Piezoelectric inkjet (PIJ) is a vital technology that was firstly used in press industry, and was later introduced to a diversity of newly-emerging applications. Due to its ability to precisely deposit a wide range of different materials including polymers, metals, ceramics, etc.\citep{DeGans2004}\citep{Kamyshny2011}\citep{Derby2011}, on various substrates, it is now heavily utilized in display, electronic device, life science and 3D printing industries\citep{Dijksman2007}\citep{McKerricher2017}\citep{Saunders2014}. 
To enhance the efficiency and quality of printing, intensive investigations have been carried out to explore the influence of both the driving parameters and the fluid properties on inkjet printing. Those studies mainly focus on two highly interrelated areas, viz., the acoustic properties of the ink channel and the drop formation process. For the former, Bogy and Talke\citep{Bogy1984} first took into account the compressibility and used linear acoustics to explain the wave propagation phenomena involved in drop ejection. Their studies successfully revealed the relationships of the optimal voltage pulse width, the meniscus motion and the oscillation with the cavity length. Wijshoff\citep{Wijshoff2006} gained insight into more complicated factors of acoustics like ‘cross talk’ effect and put forward several drop-size-modulation (DSM) techniques to fire favorable droplets. For the latter, namely, the drop dynamics, the theoretical foundation dated back to Rayleigh’s works\citep{Rayleigh1879} on instability of jets.One of the most knotty and critical problem in PIJ is the pinch-off mechanism because the breakup of the tail behind the primary drop can lead to satellites which would deteriorate the printing quality. To guarantee the printing stability, the ratio of Reynolds number to Weber number was first proposed as a criterion by Fromm\citep{Fromm1984} , and was then widely accepted. Zhang et al.\citep{Zhang1999}  elucidated the effects of ink properties on the dynamics of drop formation through experiments and simulations. Zhong et al.\citep{Zhong2018}  considered both the driving parameters of drop-on-demand (DOD) PIJ and the liquid properties, then proposed a new non-dimensional criterion \textit{Pj} with which prior determination for stable inkjet process is better made.

So far the studies scarcely paid attention to the motion states of a single stable droplet after the pinch-off, whereas it cannot be neglected that the shape and velocity of the moving drop can significantly influence the deposition\citep{Yun2017}\citep{Rein1993}. Earlier work by Reis\citep{Reis2005} illustrated influences of the operating parameters on the volume and velocity of particle suspensions. Shin\citep{Shin2015} did similar research using aerosol particles. As for the droplet oscillation, Rayleigh\citep{Rayleigh1879} was the first who investigated the free oscillation of a drop through mathematical deduction and obtained an ideal linear relationship between frequency and $\sqrt{\sigma /\rho {{r}^{3}}}$. Lamb\citep{Lamb1932} studied irrotational flow, incompressible fluids with small and axisymmetric deformations with a low viscosity. Becker\citep{Becker1991} et al. examined experimentally and theoretically the nonlinear effects involved in large-amplitude oscillation. U Kornek\citep{Kornek2010} used soap bubbles to test the theoretical models set by predecessors in this field. Despite that these studies gained deep understanding of the physics behind the oscillation phenomenon, the bridge between the oscillation theory and applications in PIJ remains unclearly.

In this study we focus on the deformation of single stable droplets exited by DOD piezoelectric inkjet system. The study includes two parts, both of which are treated by experiments and simulations. In the first part, changes of velocity and volume are investigated and explained based on Bogy’s theory\citep{Bogy1984}. In the second part, the oscillation of the droplets are observed and described.

\section{Experimental description}

Experiments were carried upon a DOD piezoelectric inkjet system that consists of two parts: the droplet generation module and the visual observation module. The droplet generation module comprises a pressure controller, a reservoir, drive electronics and a MJ-AT-01 piezo nozzle with an orifice radius of 40 $\mu m$. The pressure controller can precisely regulate the back pressure of the system. The driving electronics is connected to the PC, and offers different forms of drive pulses to the piezo-nozzle. The visual observation module comprises a group of displacement benches, a LED strobe, and a CCD camera equipped with a four times magnifier. The LED strobe produces flashes with an exposure time less than 10 $\mu s$ to prevent ‘smearing’ of the drop during its journey. The CCD camera has a frame rate of 22 fps, and an effective pixel of 1280 $\times $ 960. A single pixel size is 3.75 $\mu m$, so the actual size of the drop can be calculated by multiplying the measured size of 3.75$/$4.

\begin{figure}
	\centerline{\includegraphics[width=0.9\textwidth]{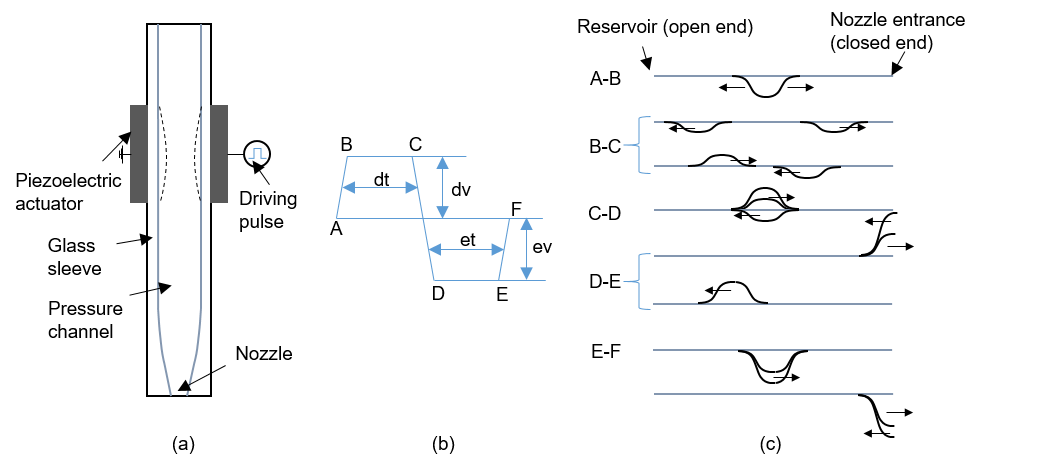}}
	\caption{(\textit{a}) Structure of the piezo-nozzle, (\textit{b}) the driving pulse, and (\textit{c}) propagation of pressure wave in the pressure channel according to linear acoustics.}
	\label{fig2}
\end{figure}

Figure \ref{fig2} shows the detailed structure of the piezo-nozzle, the drive pulse and the resulting channel acoustics. We use a bipolar drive pulse that has four important parameters: dwell voltage (\textit{dv}), dwell time (\textit{dt}), echo voltage (\textit{ev}) and echo time (\textit{et}). In our study, dwell voltage of 44 \textit{V} -85 \textit{V}, dwell time of 37 $\mu s$ -41 $\mu s$, echo voltage of 32 \textit{V} -48 \textit{V}, and echo time of 32 $\mu s$-50 $\mu s$ are used. After the pulse is applied, a deformation of the PZT tube together with the channel occurs accordingly. As shown in fig. \ref{fig2} (\textit{b}) and 2 (\textit{c}), the first slope of the driving form (A-B) leads to an enlargement of the channel, and the resulting negative pressure propagates in opposite directions with half of the amplitude. The pressure is reflected with a reverse phase at the reservoir and with the same phase at the nozzle entrance (B-C). The second slope (C-D) reduces the channel to a size smaller than its original, adding amplitude to the positive pressure traveling to the right. The compound positive pressure arriving at the nozzle entrance pushes the ink out and finally causes the drop formation (D-E). The negative pressure traveling to the left is eliminated, and a positive pressure is generated due to the over-contraction. The residual pressure in the channel will continue to propagate and decay. The third slope restores the channel to its original size, and a negative pressure is generated. As the negative pressure reached the nozzle, it pulls back the fluid column, though its effect is limited.

\section{Mathematical model}
In numerical simulations, the transient motion of the liquid is governed by incompressible Navier-Stokes equations,

\begin{equation}
\rho (\frac{\partial \mathbf{u}}{\partial t}+\mathbf{u}\cdot \nabla \mathbf{u})-\mu \Delta \mathbf{u}+\nabla p={{\mathbf{F}}_{st}}
\end{equation}

\begin{equation}
(\nabla \cdot \mathbf{u})=0
\end{equation}

where $\rho$ denotes density, $\mu$ denotes the dynamic viscosity, $\textbf{u}$ is the velocity, p is the pressure, and ${\mathbf{F}}_{st}$ is the surface tension. The surface tension is dependent on the interface normal n of the liquid. To trace the shape and place of the interface, a level set method is adopted. The convection of the reinitialized level set function is as follows\citep{Olsson2007}\citep{Olsson2005}:

\begin{equation}
\frac{\partial \Phi }{\partial t}+\nabla \cdot (\Phi \mathbf{u})+\gamma [(\nabla \cdot (\Phi (1-\Phi )\frac{\nabla \Phi }{|\nabla \Phi |}))-\varepsilon \nabla \cdot \nabla \Phi ]=0
\end{equation}

Here $\Phi$ is the level set function, $\varepsilon$ determines the thickness of the transition layer and $\gamma$ determines the amount of reinitialization. The interface is defined as the 0.5 contour of $\Phi$, where $\Phi$ equals 0 in air and 1 in liquid. The surface tension can be computed by

\begin{equation}
{{\mathbf{F}}_{st}}=\nabla \cdot \mathbf{T}
\end{equation}

\begin{equation}
\mathbf{T}=\sigma (\mathbf{I}-(\mathbf{n}{{\mathbf{n}}^{\mathbf{T}}}))\delta
\end{equation}

where \textbf{I} is the identity matrix, and $\sigma$ denotes the surface tension coefficient. $\delta$ is a Dirac delta function that is nonzero only at the fluid interface approximated by

\begin{equation}
\delta =6|\Phi (1-\Phi )||\nabla \Phi |
\end{equation}
An axisymmetric 2D model is utilized. The zone between the inlet and the nozzle is filled with liquid initially while the zone from the nozzle to the substrate is filled with air. Then the velocity boundary condition at the inlet is set as

\begin{equation}
v(r)=4{{V}_{in}}(\frac{r+0.1mm}{0.2mm})(1-\frac{r+0.1mm}{0.2mm})m/s
\end{equation}

where ${{V}_{in}}$ denotes the given velocity in the middle of the parabolic profile, and \textit{r} is the radial position. To imitate the actual velocity at the nozzle where there is a push-out and a draw-back process, we use a smooth step function to change the velocity with time, so the time-dependent velocity is

\begin{eqnarray}
v(r,t)=&v(r)\cdot\{[step(t-1\cdot 10^{-6})-step(t-(t_{p}+1\cdot10^{-6}))]\nonumber \\-&1.2[step(t-(t_{p}+1\cdot 10^{-6}))-step(t-(t_{p}+5\cdot 10^{-6}))]\}
\end{eqnarray}

where $t_{p}$ is the time for which the push-out process lasts while the draw-back time is set as a constant of 4 $\mu s$.
The outlet is set at a constant pressure. No slip condition is set on all walls except the substrate where a wetted wall with a contact angle of $\pi/2$ and a slip length of 10 $\mu m$ are adopted. To guarantee that a single stable drop without satellites could be generated in the cases, we used an inlet velocity of 0.7-1.025 \textit{m/s}, a push-out time of 13-25 $\mu s$, a density of 600-850 $kg/m^{3}$, a viscosity of 0.5-6 \textit{cp}, and a surface tension coefficient of 0.02-0.06 \textit{N/m}.

A mesh-independent test has been carried out before formal calculations so that the simulation results are not sensitive to the grids. A mesh system of 30646 grids is employed with an adaptive meshing technique to refine the region around the interface.

\section{Results and discussion}

\subsection{Velocity and equivalent radius of the droplets}

\begin{figure}
	\centerline{\includegraphics[width=0.9\textwidth]{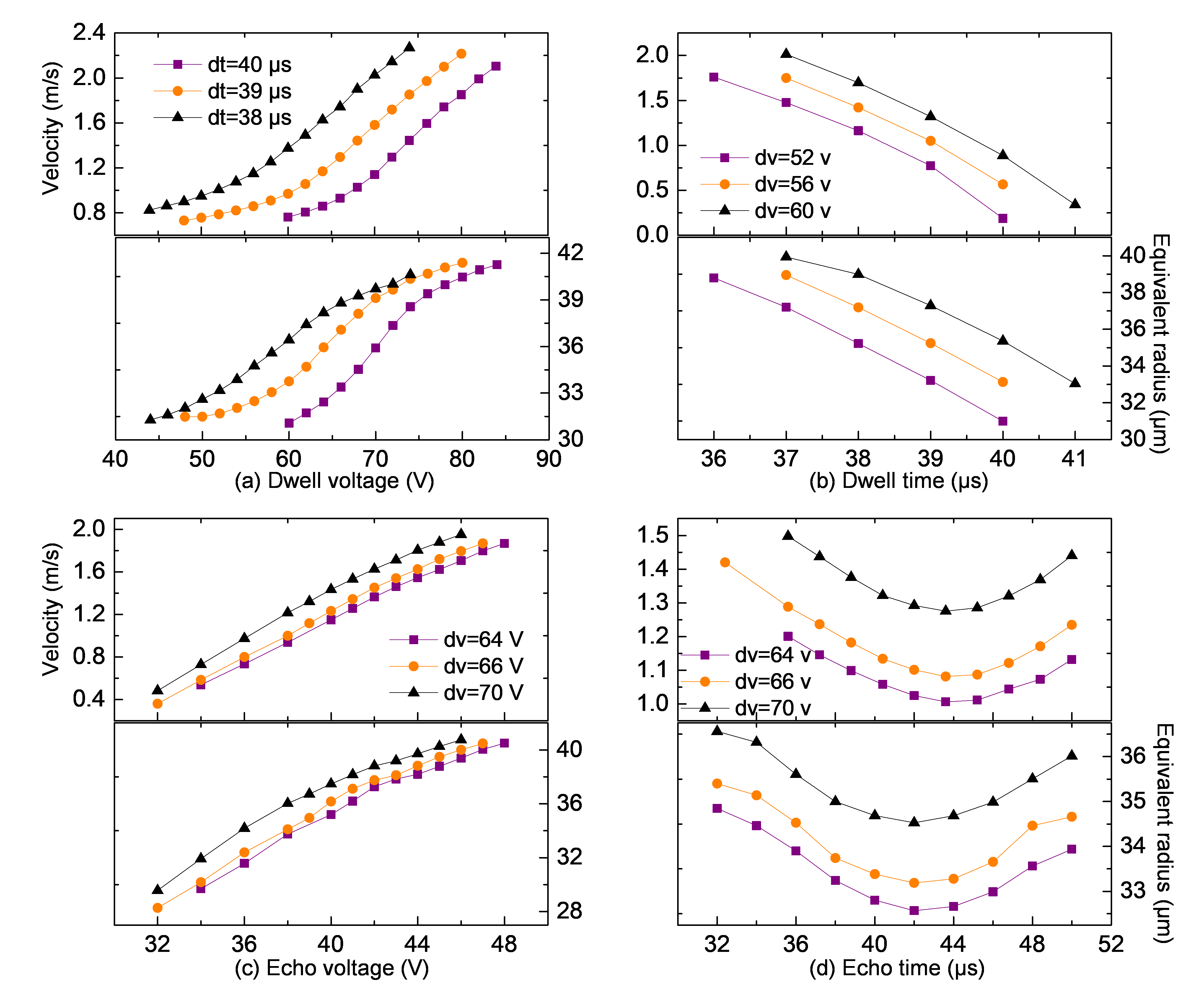}}
	\caption{Velocity and equivalent radius as a function of (\textit{a}) the dwell voltage, (\textit{b}) the dwell time, (\textit{c}) the echo voltage, and (\textit{d}) the echo time.}
	\label{fig4}
\end{figure}

In this section, we explore how the velocity and equivalent radius are influenced by operating parameters and fluid properties. As shown in Figure \ref{fig4} (\textit{a}), the velocity and equivalent radius can be adjusted smoothly and finely by changing the dwell voltage. Three groups of experiments with varying dwell times were made, and the results indicates a same trend. When applied a higher dwell voltage, the speed rises almost linearly when the velocity is above 1.2 \textit{m/s}. The equivalent radius of the drop also increases with rising dwell voltage. Its growth rate first increases and then decrease. The effects of the dwell time is illustrated in fig. \ref{fig4} (\textit{b}). It can be observed that the velocity and volume of the drop can be largely reduced as dwell time rises. By raising the echo voltage one can accelerate the drop and increase its volume as well, illustrated in fig. \ref{fig4} (\textit{c}). The underlying mechanism is the same as that of the dwell voltage because a rise of either dwell voltage or echo voltage can magnify the deformation of piezoelectric tube, thus enlarging the driving pressure that pushed the liquid out of the nozzle. The influence of echo time is relatively slight compared with the other three parameters. The speed and equivalent radius of the ejected drops first decline, and then go up with increasing echo time.

\begin{figure}
	\centerline{\includegraphics[width=0.8\textwidth]{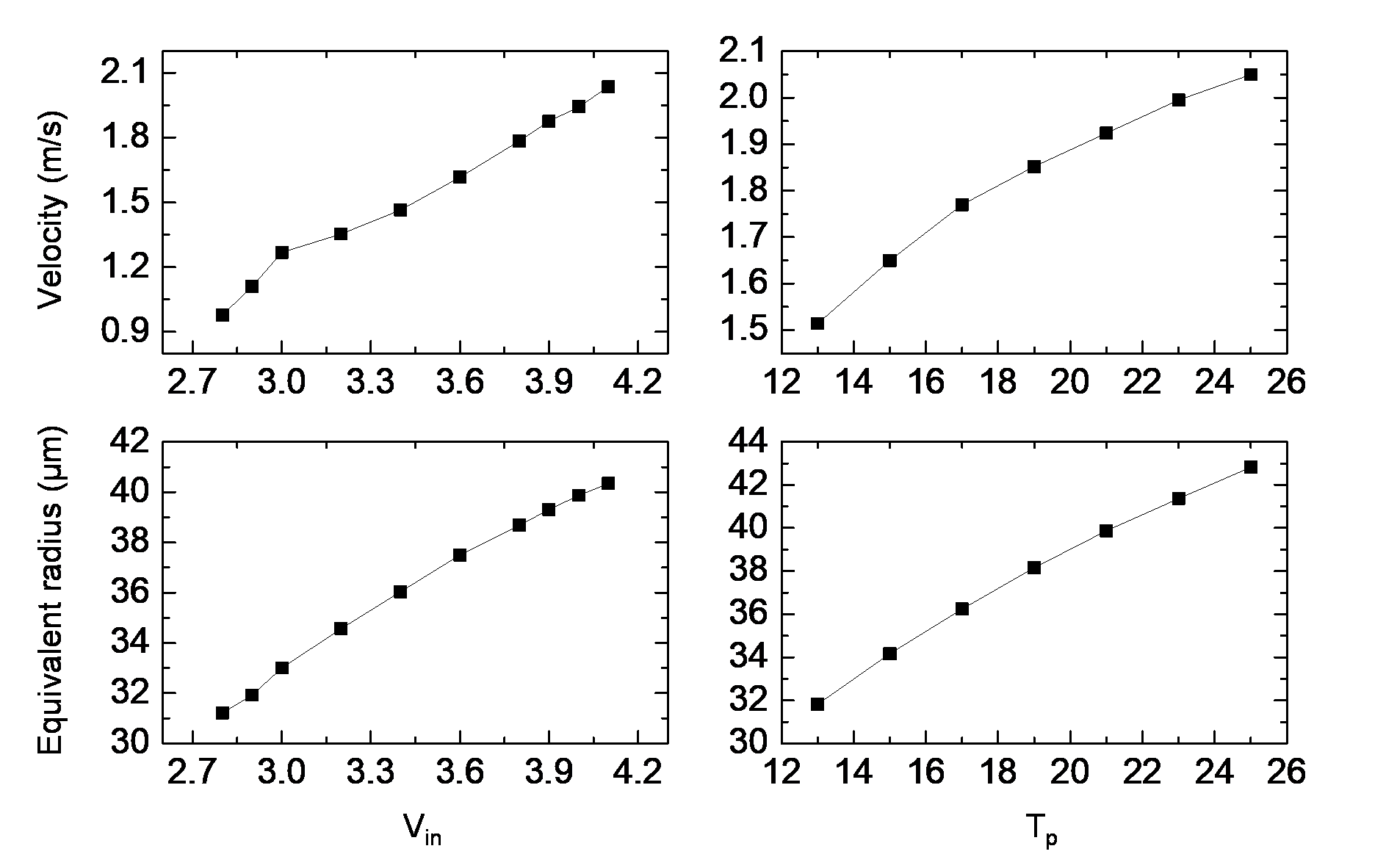}}
	\caption{Influence of ${{V}_{in}}$ and $t_{p}$.}
	\label{fig5}
\end{figure}

To explain these trends we can attribute the resulting speed and radius to two parameters in simulation: inlet velocity ${V}_{in}$ and push-out time ${t}_{p}$, as shown in Figure \ref{fig5}. Rise of dwell voltage and echo voltage both raise the amplitude of pressure thus increasing the speed and volume of the droplets. As for the effect of dwell time, it is a more complicated problem because it affects two parameters. When the dwell time rises the width of the actual wave expands (equivalent to an increase of ${t}_{p}$ in simulation) while its amplitude descends (equivalent to a decrease of ${V}_{in}$ in simulation). A larger ${t}_{p}$ will produce droplets with higher velocity and bigger size while a less ${V}_{in}$ will produce droplets with lower velocity and smaller size. As a result, the droplets velocity and size are determined by a combination of these two factors. A comparison of experimental and simulated results reveals that between these two factors the reduction of ${V}_{in}$ plays the overriding role since the speed and equivalent radius of the drop almost diminish linearly as the dwell voltage increases. Similarly the influence of the echo time is the co-effect of the pull-back amplitude and time determined by the reflected negative wave initiated by contraction (fig. \ref{fig2}(\textit{b}) C-D) and the negative pressure produced by the restoration of the cavity (fig. \ref{fig2}(\textit{b}) E-F). 

\begin{figure}
	\centerline{\includegraphics[width=1.2\textwidth]{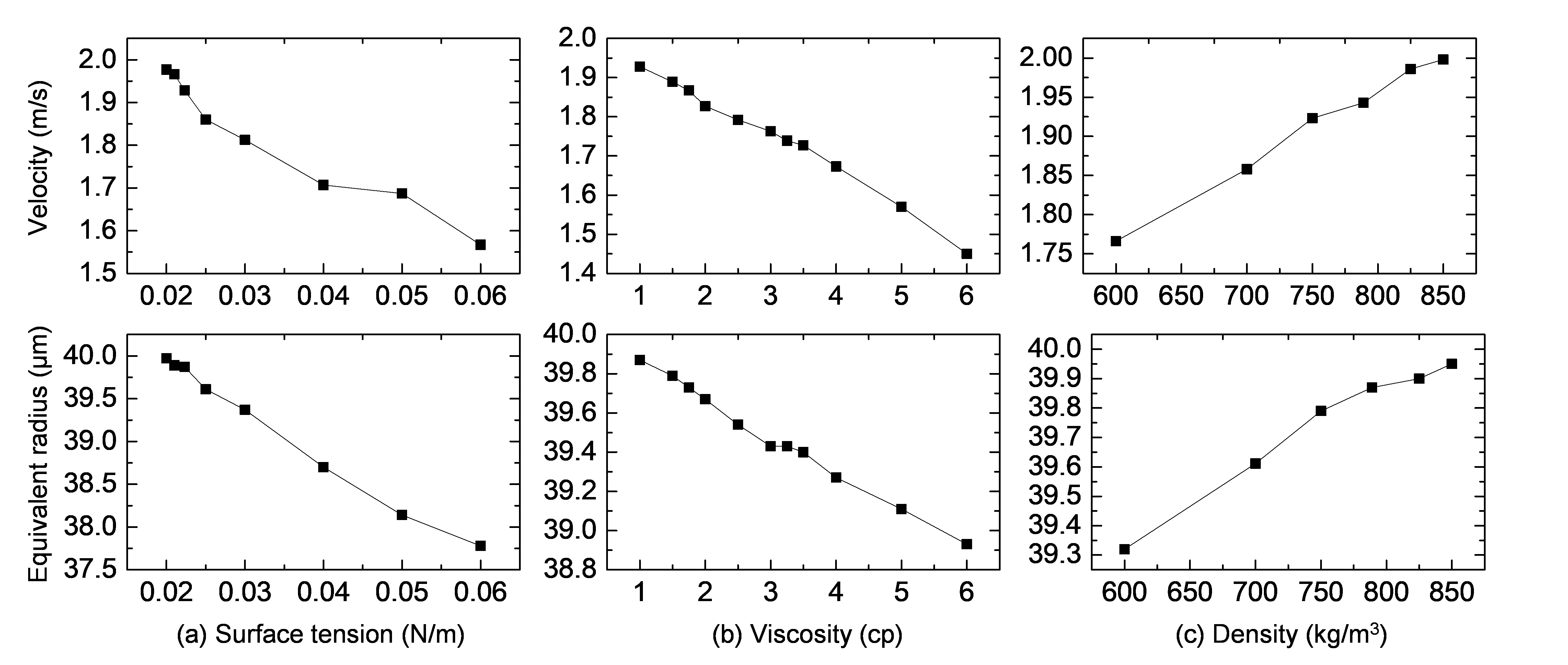}}
	\caption{Influence of (\textit{a}) surface tension, (\textit{b}) viscosity and (\textit{c}) density on drop velocity and radius.}
	\label{fig6}
\end{figure}

The influences of fluid properties, that is, surface tension, viscosity and density, are simulated and illustrated in Figure \ref{fig6}. Larger surface tension coefficient and viscosity cause larger surface tension force to overcome and greater viscosity loss at the nozzle. Thus, with the same driving force (${V}_{in}$) and time (${t}_{p}$), the velocity and equivalent radius of the produced droplets decrease as surface tension and viscosity increase. Density has relatively negligible influence on the volume of drop because in this study we set the velocity distribution as boundary condition so that as long as the operating time of ${V}_{in}$ is the same the volume of the liquid column squeezed out will be of approximately the same.

\subsection{Oscillation of the droplets}

\begin{figure}
	\centerline{\includegraphics[width=1.1\textwidth]{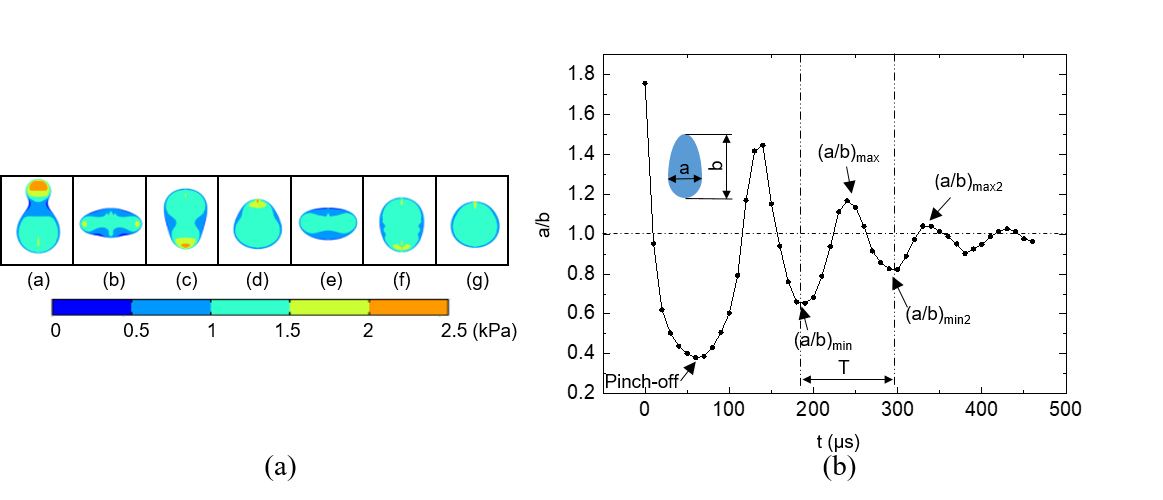}}
	\caption{(\textit{a}) Pressure distribution of the falling droplet, and (\textit{b}) deformation of the droplet in the air under condition of \textit{dv}=52 \textit{V}, \textit{dt}=38 $\mu s$, \textit{ev}=42 \textit{V} and et=40 $\mu s$.}
	\label{fig7}
\end{figure}

Figure \ref{fig7} (\textit{a}) illustrates a typical sequence of surface profiles and pressure distribution inside the drop. Uneven distributed pressure resulting from different surface tension according to Young-Laplace equation accelerated or decelerated parts of the droplets, leading to similar circles of deformation. Figure \ref{fig7} (\textit{b}) shows the oscillation of droplets in the air after they are ejected from the nozzle. The deformation is characterized by two parameters: the ratio of the width of the droplet to its length (a/b), and the oscillation period between two analogous forms of the drop (T). At ${t}_{0}$ the droplet transforms to an elongated ellipse with a minimum (a/b) for the first time, and the oscillation circle begin. We regard this point as the ${(a/b)}_{min}$. After a period of time, (a/b) reaches its local maximum marked as ${(a/b)}_{max}$ point. Then, at ${t}_{0}$+T, (a/b) reaches again a local minimum defined as ${(a/b)}_{min2}$. We define other local maxima and minima points in this way.

\begin{figure}
	\centerline{\includegraphics[width=0.9\textwidth]{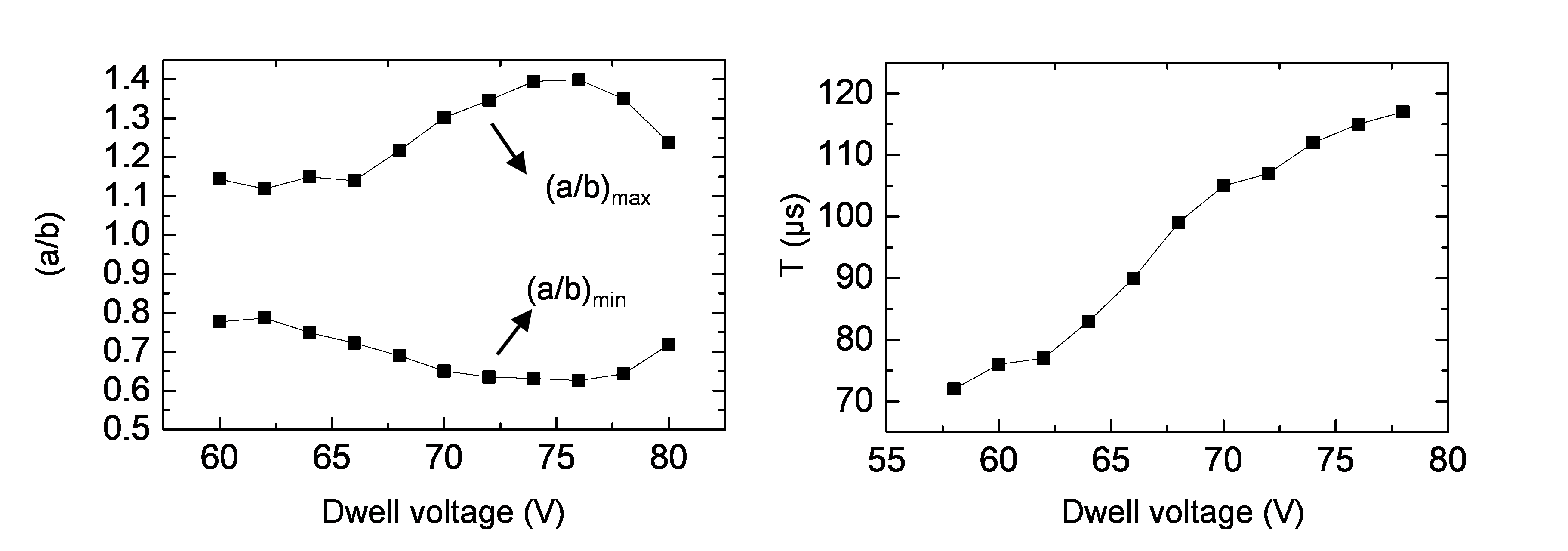}}
	\caption{Influence of dwell voltage (\textit{a}) on the ratio a/b, and (\textit{b}) on the deformation cycle T.}
	\label{fig9}
\end{figure}

Here we mainly focus on ${(a/b)}_{max}$, ${(a/b)}_{min}$ and T. From Figure \ref{fig9} it is noted that ${(a/b)}_{max}$ first increases and then decreases as the dwell voltage increases, while ${(a/b)}_{min}$ undergoes an opposite trend against ${(a/b)}_{max}$. For convenience we use $\varepsilon$=${(a/b)}_{max}$ -1 to represent the maximum deformation extent of a droplet. Then we can see that the droplet deformation extent has a maximum value of approximate 0.4 for a dwell voltage of 75\textit{V} at the specific working condition for \textit{dt}=39 $\mu s$, \textit{ev}=40 \textit{V} and \textit{et}=40 $\mu s$. From the experiment we find that the oscillation is almost cyclic, and thus we use the time interval between ${(a/b)}_{min}$ and ${(a/b)}_{min2}$ as the deformation cycle. From fig. \ref{fig9} we can see that when the dwell voltage is set to higher values, the deformation cycle will extend accordingly.

\begin{figure}
	\centerline{\includegraphics[width=0.9\textwidth]{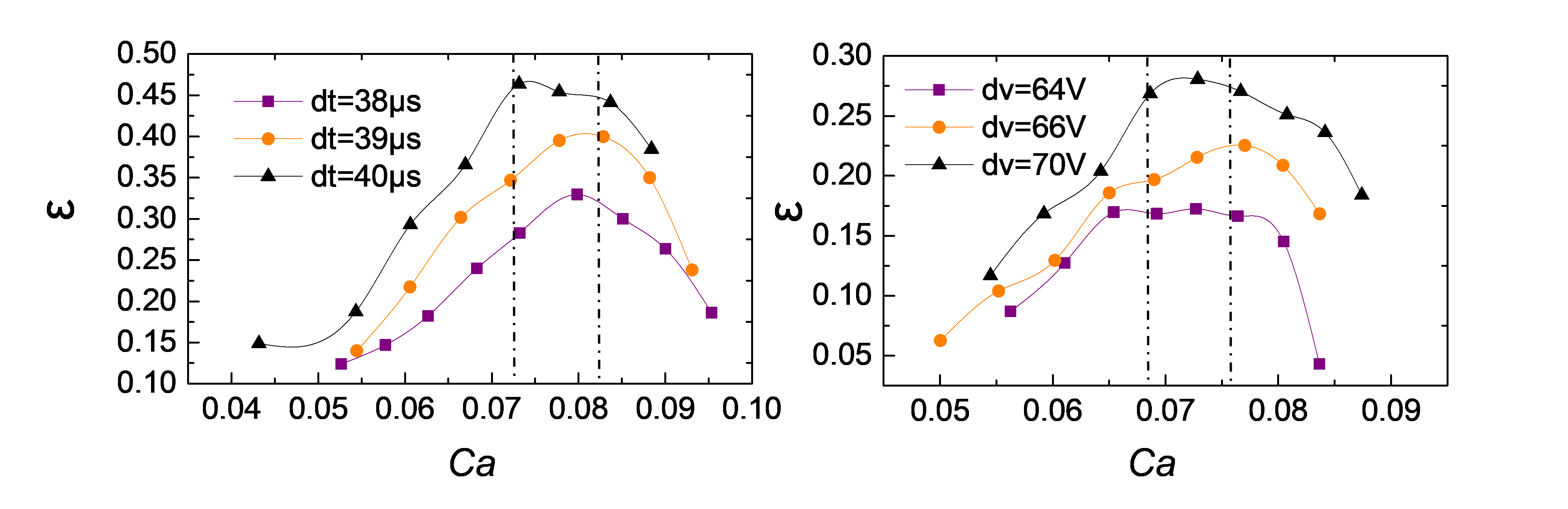}}
	\caption{Deformation extent $\varepsilon$ versus \textit{Ca} (\textit{a}) by varying the dwell voltage at three different dwell times, and (\textit{b}) by varying the echo voltage at three different dwell voltages.}
	\label{fig10}
\end{figure}

The deformation extent of droplets is mainly affected by two liquid properties: viscosity and surface tension. Therefore, we may intuitively relate $\varepsilon$ with the capillary number \textit{Ca}, which describes the relative effect of viscous drag force and surface tension force.

\begin{equation}
Ca=\frac{v\mu }{\sigma }
\end{equation}

By manipulating operating parameters, we change the velocity of the droplet. As the velocity increases, the capillary number increases accordingly. What is interesting about the oscillation is that the deformation extent does not go through a monotonous increase with \textit{Ca} but has a maximum value within the range of 0.07-0.082. Moreover, at different dwell times or dwell voltages, the deformation extent has different values even \textit{Ca} is the same. This illustrates the limitations of using the conventional criteria for determining the drop oscillation process as it is also influenced by the operating conditions, which has a great impact on the ejection period and the initial disturbance of the droplets.

\begin{figure}
	\centerline{\includegraphics[width=0.9\textwidth]{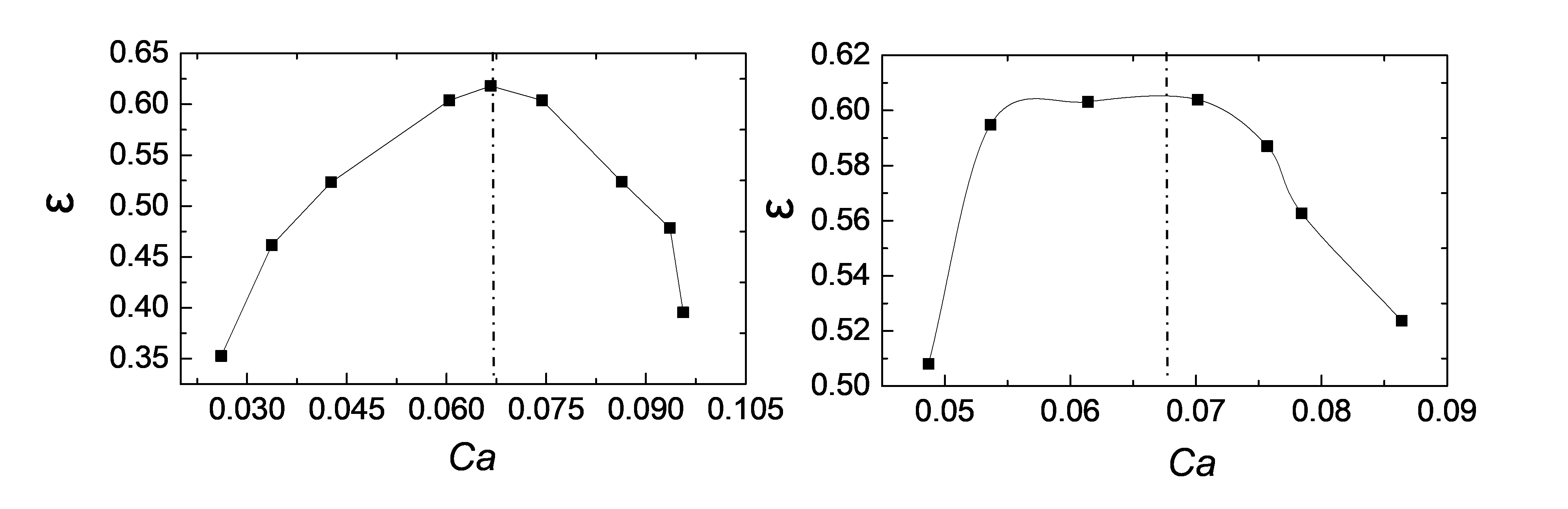}}
	\caption{Deformation $\varepsilon$ versus \textit{Ca} by varying (\textit{a}) surface tension, and (\textit{b}) viscosity.}
	\label{fig11}
\end{figure}

\begin{figure}
	\centerline{\includegraphics[width=0.55\textwidth]{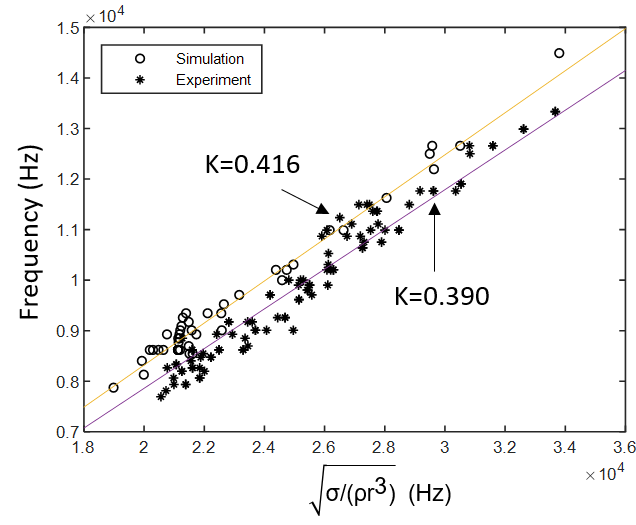}}
	\caption{Frequency as a function of $\sqrt{\sigma /\rho {{r}^{3}}}$.}
	\label{fig12}
\end{figure}

In numerical simulation we change \textit{Ca} by varying the liquid properties while keeping the driving pulse constant. As a consequence, the velocity of the ejected droplet will change accordingly. Therefore, the value of \textit{Ca} is determined by the resulting velocity and surface tension or viscosity simultaneously. However, it is found that as \textit{Ca} increases the deformation extent represents the same trend, and reaches its maximum for \textit{Ca} at around 0.68, slightly less than the lower limit of \textit{Ca} in our experiment. From the discussion above, we find that there is always a critical \textit{Ca} where the deformation extent reaches the maximum value under the same operating parameters. The critical \textit{Ca} falls within the range of 0.68-0.82.

Through dimensional analysis we find the oscillation frequency \textit{f}, the reciprocal of deformation cycle, has a linear correlation with $\sqrt{\sigma /\rho {{r}^{3}}}$:

\begin{equation}
f=k\sqrt{\sigma /\rho {{r}^{3}}}
\end{equation}

From Figure \ref{fig12}, it is shown that all the data points collapse on the straight lines. The slope of the lines k are 0.416 and 0.390 respectively for simulation and experiment, both of which are smaller than Rayleigh’s\citep{Rayleigh1879} mathematical prediction $\sqrt{2}/\pi$ .

\section{Conclusions}
Drop-on-demand inkjet technology has been of great scientific interest, and has been used in areas of conventional printing, mechanical engineering, electronics and biochemistry. Given the crucial role of the droplet velocity, the droplet diameter and the droplet oscillation in precise deposition process, all of them are investigated with single droplets. By means of numerical and experimental method, the influences of driving parameters including dwell voltage (\textit{dv}), dwell time (\textit{dt}), echo voltage (\textit{ev}) and echo time (\textit{et}) as well as ink properties including density, viscosity and surface tension are presented. Results show that the rise of the dwell voltage and the echo voltage can effectively increase the velocity and radius of droplets, while the dwell time and the echo time have complex impacts on droplets’ motion due to the complexity of pressure wave propagation within ink channel. A higher surface tension, viscosity or a lower density value can considerably reduce the velocity of the droplet at the same operating condition, whereas the volume of the drop is mainly related to the surface tension. Through analyzing characteristics including deformation extent and deformation cycle during the process of droplet oscillation, it is found that deformation is inherently driven by the shift of pressure distribution inside the droplet after its tail merges into its head. As driving parameters increase the deformation extent first increases then decreases. As for fluid properties, viscosity influences the deformation extent most and surface tension influences the deformation cycle most. The capillary number (\textit{Ca}) is found to be related with the deformation extent, which reaches its maximum value when \textit{Ca} locates within the range of 0.68-0.82. Additionally, the oscillation frequency of ejected droplet is found to be linearly related with $\sqrt{\sigma /\rho {{r}^{3}}}$ with a coefficient of approximate 0.4, which is slightly lower than Rayleigh’s mathematical prediction.

\bibliographystyle{jfm}
\bibliography{Droplet_Citation}
\end{document}